\documentclass[apj]{emulateapj}
\usepackage{natbib}

\shorttitle{Anisotropy of X-ray bursts}
\shortauthors{He \& Keek}

\newcommand{\slugcom}{Submitted for publication in The Astrophysical Journal}
\slugcomment{\slugcom}

\begin{document}

\title{Anisotropy of X-ray bursts from neutron stars with concave accretion disks}

\author{C.-C. He\altaffilmark{1}}
\affil{College of Physics, Jilin University, Changchun 130012, China}
\email{jordanhe1994@gmail.com}

\and

\author{L. Keek}
\affil{CRESST and X-ray Astrophysics Laboratory NASA/GSFC, Greenbelt, MD
20771}
\affil{Department of Astronomy, University of Maryland, College Park, MD
20742}

\altaffiltext{1}{Undergraduate, Tang Aoqing Honors Program in Science}

\begin{abstract}
Emission from neutron stars and accretion disks in low-mass X-ray
binaries is not isotropic. The non-spherical shape of the disk as
well as blocking of the neutron star by the disk and vice versa cause
the observed flux to depend on the inclination angle of the disk with
respect to the line of sight. This is of special importance for the
interpretation of Type I X-ray bursts, which are powered by the thermonuclear
burning of matter accreted onto the neutron star. Because part of
the X-ray burst is reflected off the disk, the observed burst flux
depends on the anisotropies for both direct emission from the neutron
star and reflection off the disk. This influences measurements of
source distance, mass accretion rate, and constraints on the neutron
star equation of state. Previous studies made predictions of the anisotropy
factor for the total burst flux, assuming a geometrically flat disk.
Recently, detailed observations of two exceptionally long bursts (so-called
superbursts) allowed for the first time for the direct and the reflected
burst flux to each be measured, as opposed to just their sum. The
ratio of the reflected and direct flux (the reflection fraction) was
much higher than what the anisotropies of a flat disk can account
for. We create numerical models to calculate the anisotropy factors
for different disk shapes, including concave disks. We present the
anisotropy factors of the direct and reflected burst flux separately,
as well as the anisotropy of the persistent flux. Reflection fractions
substantially larger than unity are produced in case the inner accretion
disk steeply increases in height, such that part of the star is blocked
from view. Such a geometry could possibly be induced by the X-ray
burst, if X-ray heating causes the inner disk to puff up.
\end{abstract}

\keywords{accretion, accretion disks --- stars: neutron --- X-rays: binaries --- X-rays: bursts}

\section{Introduction}

\label{sec:intro}

Thermonuclear burning on the surface of an accreting neutron star
in a low-mass X-ray binary (LMXB) is observed to produce Type
I X-ray bursts \citep{Grindlay1976,1976Belian,Lewin1993,Galloway2008catalog}.
Hydrogen and/or helium-rich material is accreted through a disk from
the companion star, and collects on the neutron star surface until
runaway thermonuclear burning ignites \citep{Woosley1976,Maraschi1977,Lamb1978}.
As the thermonuclear burning rate increases dramatically, the surface
layers burn on a timescale of seconds, powering an X-ray burst with
a typical duration of $10$ to $100\,\mathrm{s}$. The heated photosphere
thermally emits a spectrum close to a blackbody \citep{swank1977,Suleimanov2010}.
The burst flux surpasses the persistent X-ray flux originating from
the inner accretion disk. Type I X-ray bursts are employed as standard
candles for distance determination \citep[e.g.,][]{Kuulkers2003},
to study the nuclear physics of proton-rich unstable isotopes \citep[e.g.,][]{Schatz2006},
and to constrain the equation of state of dense matter by measuring
the neutron star mass and radius \citep[e.g.,][]{Ozel2006,Suleimanov2011}. 

For all use cases, it is important that an accurate measure of the
neutron star's luminosity is derived from the observed flux. This
poses two challenges. First, the neutron star flux must be distinguished
from the persistent X-ray flux as well as from ``reflection'': the
diffuse scattering of the neutron star emission off the disk. Because
X-ray bursts typically have short durations, the spectra are of insufficient
quality to separate direct thermal emission from the neutron star
from other components. An often used approach is to observe the persistent
spectrum outside the burst, and assume that it remains constant during
the burst. This works well in practice, as the burst emission is much
stronger than the persistent emission, at least around the burst peak
\citep[see also the discussion in][]{2002Kuulkers}. Recently, however,
observations have shown that the persistent component does evolve
during the burst \citep{Worpel2013,Zand2013,Ji2014,Keek2014sb1}.
Furthermore, as the reflection spectrum resembles that of the neutron
star flux \citep{Ballantyne2004models}, they are typically not distinguished.
The observed burst flux is then a combination of the direct and reflected
burst emission. Only in the case of two superbursts, could the Fe~K$\alpha$
line from reflection be detected, and could the reflection component
be separated from the direct burst component \citep{Ballantyne2004,Keek2014sb1,Keek2014sb2,Keek2015sb3}.
Superbursts are exceptionally long bursts that last hours \citep{Cornelisse2000,Strohmayer2002,Kuulkers2003a,Keek2008int..work}.
They are, however, rare, and only in those two cases were relatively
high quality spectra obtained. We anticipate that future large-area
X-ray observatories will have the capability to distinguish direct
from reflected burst emission for a large sample of bursts (Wolf et
al. 2016 in prep.), such as the \emph{Neutron Star Interior Composition
Explorer} \citep[NICER,][]{Gendreau2012NICER} and \emph{Athena} \citep{Barcons2015Athena}.

Second, the flux is anisotropic: part of the line-of-sight to the
neutron star is blocked by the disk, and the degree of anisotropy
depends on the inclination of the disk. Analytic \citep{fujimoto88apj}
and numerical models \citep{Lapidus1985} have been used to investigate
the effect of anisotropy on the observed flux under the assumption
of a geometrically thin accretion disk. The degree by which the observed
flux deviates from the isotropic flux was expressed in anisotropy
factors for both the persistent and the total burst flux (direct +
reflection). The anisotropy factors are important when comparing observed
X-ray bursts to theoretical predictions \citep[e.g.,][]{Heger2007},
as they influence measurements of the distance from photospheric radius
expansion (PRE), the mass accretion rate, and the $\alpha$-parameter,
which is used to characterize the nuclear burning regime and fuel
composition \citep[e.g.,][]{Chenevez2015}. Often studies choose to
ignore the anisotropies, however, because the geometry and the inclination
angle of the disk are poorly constrained. Only when dips or eclipses
are observed, do we have indications that the inclination is large
\citep{Frank1987}.

The detection of reflection features in two superbursts provides a
new observational constraint on the anisotropy factors. The observed
ratio of the direct and reflected burst flux (the reflection fraction)
depends on the anisotropy factors, which in turn depend on the geometry
of the system. Assuming a flat disk, \citet{fujimoto88apj} predicts
a maximum observed reflection fraction of $0.5$. During the 1999
superburst from 4U~1820--30 \citep{Ballantyne2004} and the 2001
superburst from 4U~1636--536 \citep{Keek2014sb2}, however, reflection
fractions were observed of up to $3$ and $6$, respectively. This
may indicate that the accretion disk geometry was not flat during
the burst. A disk in equilibrium is expected to be thin \citep{Shakura1973}, but it has been suggested that persistent accretion requires a concave shape, as the outer disk needs irradiation by the inner part to maintain its ionization state \citep[e.g.,][]{Paradijs1996,King1998}. Even stronger deviations from a flat geometry may result from the intense irradiation by an X-ray
burst. For example, X-ray heating could cause expansion, or a disk
wind could be induced by the burst \citep{Ballantyne2005}. \citet{Keek2015sb3}
presented an alternative interpretation of the spectra of the superburst
from 4U~1636--536, which does not have large reflection fractions,
but poses other problems. It is, therefore, interesting to investigate
whether a disk geometry exists that can produce large reflection fractions.
For the case of accreting black holes, concave disks have been shown
to produce large reflection fractions \citep{Blackman1999}. Those
results are, however, not directly applicable to accreting neutron
stars, because the illuminating source is thought to be a corona above the disk,
rather than a star located in the disk.

In this paper we create numerical models to calculate the anisotropy
factors for a variety of disk shapes, including flat and concave disks.
We calculate the anisotropy factors separately for direct and reflected
burst emission, as well as the reflection fractions. Furthermore,
we discuss the effects of different assumptions on the angular distribution
of radiation emitted by the star and reflected off the disk, and the effect of
light bending in the strong gravitational potential close to the neutron
star.

\section{Methods}
\label{sec:Methods}

\begin{figure}
\centering
\includegraphics[width=1\linewidth]{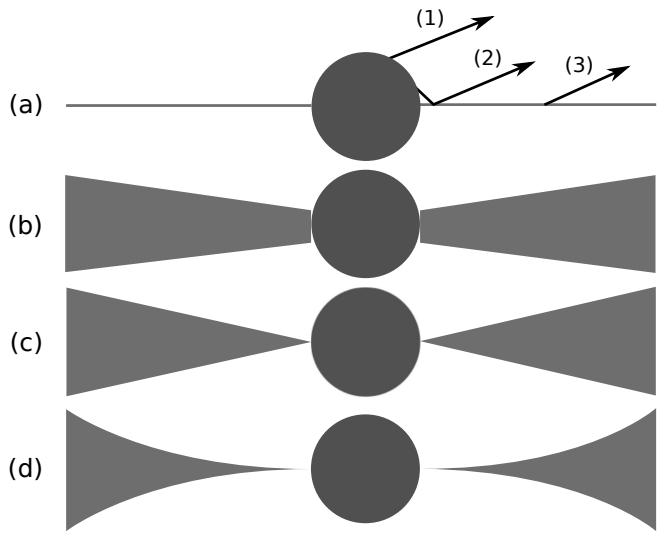}
\caption{Cartoon of four disk shapes around a star (viewed in cross section):
  flat (a), trapezoidal (b), triangular (c), and curved (d). The profile of disk
  b is such that the extrapolated height is $0$ at the center of the star, whereas for
  disks c and d it is $0$ at the surface of the star. Disk b is assumed to absorb
  all photons received on the inner side. At the top, the arrows represent three
  flux components: direct burst flux (1), reflected burst flux (2), and persistent flux (3).
  \label{fig:disk}}
\end{figure}

We first rederive the simple analytic model presented by \cite{fujimoto88apj}, which describes a thin flat disk.
Next, we create numerical models of flat as well as concave disks (Figure \ref{fig:disk}). 
When accretion disks undergo sudden strong irradiation by an X-ray
burst, their geometry may for a brief period deviate from the shape
predicted for disks in equilibrium \citep{Ballantyne2005}. We, therefore,
choose generic height profiles as disk shapes, rather than shapes
based on accretion disk theory \citep[e.g.,][]{Shakura1973}. 
The accretion environment of a neutron star may be more complex. For
example, a boundary or spreading layer may be present between the
inner disk and the star \citep{Shakura1973,Revnivtsev2001}. Depending
on the accretion flow, only part of the star may be covered, but during
X-ray bursts the entire stellar surface is thought to be covered by
the spreading layer \citep{Lapidus1985}. Our models are, therefore,
equivalent to the latter case.

Our numerical models
include realistic distributions of radiant intensity and blocking of 
the line of sight. We test the accuracy of the numerical models using the analytic model.
Before describing the models, we introduce different distribution laws of radiant intensity emitted by or reflected off plane-parallel atmospheres. 

\subsection{Angular Distribution of Radiant Intensity}
\label{sec:Radia}

\begin{figure}
\centering
\includegraphics[width=1\linewidth]{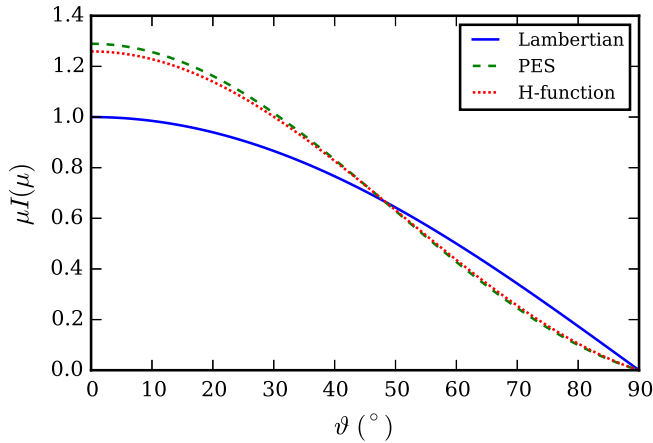}
\caption{Angular distribution of emergent radiant intensity from different atmospheres. 
The intensity is normalised such that $ \int_{0}^{\pi/2} {\rm d}\theta_0 \int_{0}^{2\pi} {\rm d}\phi ~ \mu I(\mu) \sin \theta_0 = \int_{0}^{\pi/2} {\rm d}\theta_0 \int_{0}^{2\pi} {\rm d}\phi ~\mu \sin \theta_0 = \pi $.
\label{fig:radia_transfer}}
\end{figure}


Lambert's emission law states that the intensity emitted from an ideal diffuse radiator is proportional to the cosine of the angle $ \vartheta $ between the direction of emission and the surface normal, i.e. $ I\left(\mu \right) = \mathrm{constant} $, where $ \mu = \cos\vartheta $. The intensity $ I(\mu) $ is defined such that the energy transported across an area $\mathrm{d}\sigma$ in directions confined to solid angle $\mathrm{d}\omega$ during a time $\mathrm{d}t$ is $ {\rm d}E=\mu I \left(\mu\right) \mathrm{d}\sigma \mathrm{d}\omega\mathrm{d}t $. Therefore, the flux observed under an angle $\vartheta$ is proportional to $I(\mu)\mu$.

As an alternative to Lambert's law, we use for the neutron star a pure-electron-scattering (PES) atmosphere. The angular distribution of radiant intensity emitted from the surface is given by (\citealt{Chandrasekhar1960}, see also \citealt{Lapidus1985})
\begin{equation}
I \left(\mu \right) \propto 1 + 2.06\mu.
\end{equation}

For the accretion disk we employ an $ H $-function, which applies to isotropic scattering in semi-infinite atmospheres (\citealt{Chandrasekhar1960}, see also \citealt{Lapidus1985}).
For a surface element on the disk that is illuminated by a constant incident flux $ \pi F $, the angular distribution of the reflected light is given by
\begin{equation} \label{eqn:I_H}
I \left( \mu \right)=\frac{\sqrt{3}}{4}F H\left(\mu \right),
\end{equation}
where the $H$-function is defined through an integral: 
\begin{equation} \label{eqn:I}
\log H\left(\mu \right) = -\frac{1}{2\pi} \int_{-\infty}^{\infty} \log \left( 1-m \arctan{\frac{1}{m}} \right)  \frac{\mu}{m^2+\mu^2} \mathrm{d}m,
\end{equation}
where we assume the albedo of the disk to be $1$.

The angular distribution of emergent radiant intensity is very similar for pure-electron-scattering and $H$-function-like atmospheres: they differ by at most 6\% (Figure \ref{fig:radia_transfer}). 
Compared to Lambertian atmospheres, both produce larger intensities for $\vartheta<48^\circ$ and smaller intensities for larger $\vartheta$, such that radiation is more concentrated to the surface normal direction for these two types of atmospheres.

In our analytic model, we employ Lambert's law for emission from and reflection off the atmospheres \citep[see also][]{fujimoto88apj}. 
In our numerical models, apart from Lambert's emission law, we also apply pure-electron-scattering and an $H$-function (see also \citealt{Lapidus1985}).

\subsection{Analytic Model}
\label{sec:Ana}

\begin{figure}
\centering
\includegraphics[width=1\linewidth]{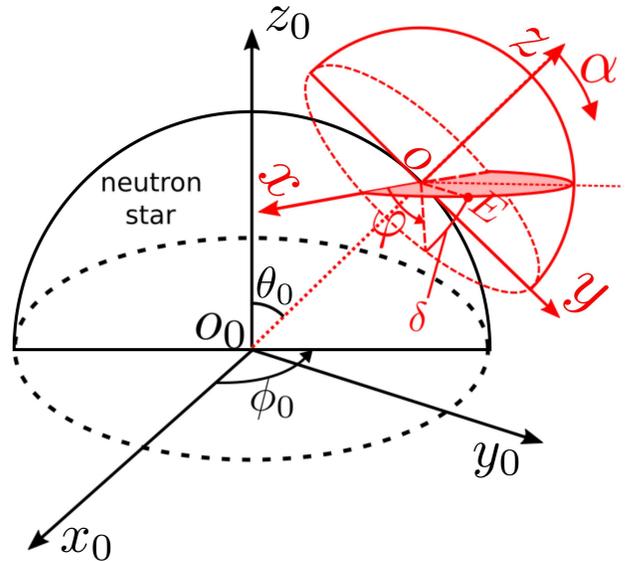}
\caption{Two spherical coordinate systems used to compute the fraction ($ P $) of photons emitted from the neutron star that irradiate the disk. The first is centered on $ o_0 $ for the neutron star, with polar angle $ \theta_0 $ and azimuthal angle $ \phi_0 $.  The second is centered on a point on the stellar surface ($ o $). For a given azimuthal angle $ \varphi $ in the $ oxyz $ coordinates, $ \delta $ is the angle such that for the polar angle $ \alpha=\pi/2-\delta $, the line $ oE $ points to the disk's outer edge, or is parallel to the disk if the disk extends to infinity.
\label{fig:coordinate}}
\end{figure}

The simple model presented by \cite{fujimoto88apj} includes a spherical non-rotating neutron star and a flat optically thick accretion disk. The inner part of the disk extends to the equator of the star and the outer part to infinity (Figure \ref{fig:disk}a). Both the stellar surface and the disk surface are assumed to be Lambertian.
The stellar surface is assumed to radiate homogeneously and isotropically. The observed burst flux, $F_\mathrm{b}$, and persistent flux, $F_\mathrm{p}$, are, however, anisotropic, and depend on the inclination angle of the disk with respect to the observer's line of sight. $F_\mathrm{b}$ includes both directly observed burst flux and burst reflection off the disk. \cite{fujimoto88apj} quantifies the deviation from isotropy with the anisotropy factors for the burst, $ \xi_{\rm b} $, and the persistent flux, $ \xi_{\rm p} $:
\begin{equation}
L_{\rm b,p}=4\pi d^2 \xi_{\rm b,p}F_{\rm b,p},\label{eq:xibp}
\end{equation}
where $d$ is the distant to the burster and $L_{\rm b,p}$ are, the burst and persistent luminosities, respectively. The anisotropy factors are normalized over solid angle: 
\begin{equation}
\frac{1}{4\pi} \oint \xi^{-1}_{\rm b,p} \mathrm{d}\omega = 1 \label{eq:norm_xis}
\end{equation}

The system is axially symmetric, and the flux measured by a distant observer depends on the inclination angle, $ \theta $, between the normal of the plane of the disk and the line of sight of the observer. As the disk is considered to be a flat Lambertian surface, the observed persistent flux is proportional to $ \cos\theta $ (Section.~\ref{sec:Radia}), so its anisotropy factor is
\begin{equation} \label{eqn:p}
\xi_{\rm p}^{-1} = 2 \left | \cos \theta \right |,
\end{equation}
where the factor $2$ is from the normalization (Equation~\ref{eq:norm_xis}).

We separately consider the direct burst flux, $F_\mathrm{d}$, and the reflected burst flux, $F_\mathrm{r}$, with $F_\mathrm{b}=F_\mathrm{d}+F_\mathrm{r}$. These flux components have different anisotropies, $\xi_\mathrm{d,r}$, which we define with respect to the total burst luminosity:
\begin{equation}
L_{\rm b}=4\pi d^2 \xi_{\rm d,r}F_{\rm d,r}.\label{eq:xidr}
\end{equation}
From Equations~\ref{eq:xibp} and \ref{eq:xidr} we see that $\xi^{-1}_\mathrm{b}=\xi^{-1}_\mathrm{d}+\xi^{-1}_\mathrm{r}$, such that the normalization of $\xi^{-1}_\mathrm{d,r}$ follows from Eq.~\ref{eq:norm_xis}.
The observed direct burst flux is proportional to the effective area because of the equal brightness effect of Lambertian surfaces. For $\theta=0^\circ$ the entire star is visible, whereas for $\theta=90^\circ$ half the star is blocked from view by the disk. This yields the anisotropy factor for the direct burst flux:
\begin{equation} \label{eqn:d}
\xi_{\rm d}^{-1} = \frac{ 1 + \left | \cos \theta \right | }{2}.
\end{equation}

$\xi^{-1}_\mathrm{r}$ includes a factor $P$, which is the fraction of $L_\mathrm{b}$ that is intercepted and subsequently scattered by the disk. We refer to $P$ as the intrinsic reflection fraction.
To compute $ P $, consider a surface element on the star with area $d \sigma = \sin\theta_0 \mathrm{d} \theta_0 \mathrm{d} \phi_0$ (Figure~\ref{fig:coordinate}). The fraction of the photons emitted from this element that irradiate the disk is given by
\begin{equation}
P(\theta_0) =\frac{1}{\pi} \int_{\frac{\pi}{2}-\delta}^{\frac{\pi}{2}} \mathrm{d} \alpha \int_{0}^{\pi} \mathrm{d} \varphi \cos \alpha \sin\alpha,
\end{equation}
with $ \delta $ as shown in Figure~\ref{fig:coordinate} (see also Equation~\ref{Eqn:tandelta} for $ a \rightarrow \infty $).
Performing the integration, we get $ P(\theta_0) = \sin^{2} \left ( \theta_0/2 \right ) $.
Integrating over one hemisphere of the star, we obtain the total intrinsic reflection fraction:
\begin{equation}
P = \frac{1}{2 \pi} {\displaystyle \int_{0}^{\frac{\pi}{2}} \int_{0}^{2 \pi} } P(\theta_0) \sin \theta_0 \mathrm{d} \theta_0 \mathrm{d} \phi_0 = \frac{1}{4}.
\end{equation}
The anisotropy factor for the reflected burst flux is similar to $\xi^{-1}_\mathrm{p}$ (Equation~\ref{eqn:p}) and includes $P$:
\begin{equation} \label{eqn:r}
\xi_{\rm r}^{-1} = 2\left | \cos \theta \right | P=\frac{\left | \cos \theta \right |}{2}.
\end{equation}
Summing Equation~\ref{eqn:d} and \ref{eqn:r}, we see that $\xi^{-1}_\mathrm{b}=\xi^{-1}_\mathrm{d}+\xi^{-1}_\mathrm{r}=1/2 + |\cos\theta|$ has the correct normalization (Equation~\ref{eq:norm_xis}), and it reproduces the result of \cite{fujimoto88apj}.

$P$ is the intrinsic reflection fraction. The observed reflection fraction is $F_\mathrm{r}/F_\mathrm{d}$.
Using Equation~\ref{eq:xidr}, we find
\begin{equation}
\frac{F_\mathrm{r}}{F_\mathrm{d}}=\frac{\xi^{-1}_\mathrm{r}}{\xi^{-1}_\mathrm{d}}.\label{eq:observed_reflfrac}
\end{equation}
For the analytic model, the observed reflection fraction is (Equations~\ref{eqn:d} and \ref{eqn:r}):
\begin{equation}
\frac{\xi^{-1}_\mathrm{r}}{\xi^{-1}_\mathrm{d}}=\frac{\left | \cos \theta \right |}{1+\left | \cos \theta \right |}.\label{eq:observed_reflfrac_analytic}
\end{equation}

In our numerical model, the disk does not extend to infinity, but ends at a radius $ r=a R_\star $,
where $ R_\star $ is the stellar radius. 
A complicated integration derived from our analytic model gives the relation between $ P_a $ and $ a $ (see Appendix \ref{app:a}). $ P_a $ for several values of $ a $ are given in Table \ref{tab:1}, which shows that most of the reflected flux originates from the inner part of the disk.

\subsection{Numerical Model} \label{sec:Num}

Similar to the analytic model, our numerical model has a spherical neutron star and an accretion disk starting from the equator of the star. 
The geometry of the disk can be flat, linearly inclined, or any concave shape (see Figure \ref{fig:disk}).
The disk extends to a finite outer radius, which we set at $r=4000\ R_\star$. Certain binaries with long orbital periods may have up to 50 times more extended disks \citep[e.g.,][]{Frank2002,Tauris2006}. We find, however, that an extent of $a=4000$ accounts for all but a fraction of $10^{-4}$ of the flux received by a flat disk that extends to infinity (Table~\ref{tab:1}).
All the simulations assume that 1) photons are emitted from the whole stellar surface isotropically and homogeneously, and 2) the accretion disk reflects all the photons that are received. Re-reflection is neglected.

As a first step, we compute the distribution of flux received by the disk as a function of the radial distance.
We equally divide the surface of the neutron star into $1024\times512$ small elements in azimuthal and polar direction. Similarly, we equally divide the surface of the accretion disk into $1024\times512$ small elements in polar direction and logarithmically in radial direction.
Elements on both surfaces are well approximated as planar. 
We take into account the reduced effective area of an element on the disk surface as viewed from an element on the stellar surface.
For each element on the disk surface we compute the flux received from each element on the stellar surface. Integrating over the whole stellar surface, we obtain the total flux received by each element on the disk. 

The second step of our model calculation is computing the angular distribution of radiation as seen by a distant observer. 
The observer is far away from this
system, and therefore the lines of sight to different points in our system are parallel. Assuming that the system is observed
under a given inclination angle $\theta$, we compute the angle 
between the observer's line of sight and the normal vectors of each element on
the disk surface. Using the flux received by each disk element, the inclination angle $\theta$, and an $H$-function, we compute the intensity reflected by each element in the direction of observer. Summation of all elements on the disk surface gives the angular distribution of intensity reflected by the whole accretion disk.

To calculate the persistent emission, we assume that the disk radiates as a blackbody. 
We take the temperature of the accretion disk to decrease with radius as \citep[e.g.,][]{Bhattacharyya2000}:
\begin{equation}
T \propto r^{-3/4} \left [ 1 - \left ( R_\star / r \right )^{1/2} \right ]^{1/4}.
\end{equation}
We compute the angular distribution of the persistent flux in the same way as the reflected burst flux, but replacing the reflected radiation profile with a blackbody radiation profile, which is 
\begin{equation}
 F \propto \sigma \ T^4.
\end{equation}
Furthermore, a pure-electron-scattering atmosphere is used instead of an $H$-function.

The angular distribution of the direct burst flux from the star is computed by considering for the visible part of the star either a Lambertian or pure-electron-scattering radiation distribution.

Observed from a certain direction, the view of the disk or the star could be (partially) blocked.
Blocking of the accretion disk by itself (disk-disk blocking), blocking of the accretion disk by the star (star-disk blocking), and blocking of the star by the disk (disk-star blocking) are taken into account.

\section{Results}

\subsection{Comparison of Numerical to Analytic Models of Flat Disks}

Our numerical model computes for Lambertian surfaces an intrinsic reflection fraction of $P=0.25010$, which has a relative difference of $0.06\%$ with the analytic model for a disk with a same radius.
Ignoring blocking of the line of sight, we find for the anisotropy factors relative differences with the analytic model of less than $0.05\%$. The numerical model is, therefore, consistent with the analytic model within $\sim 10^{-4}$. In the rest of this section, we apply the numerical model to different radiative distribution models and disk shapes.

\subsection{Radiation Models for Flat Disks}
\label{sec:radiation_flat}

\begin{figure}
\centering
\includegraphics[width=1\linewidth]{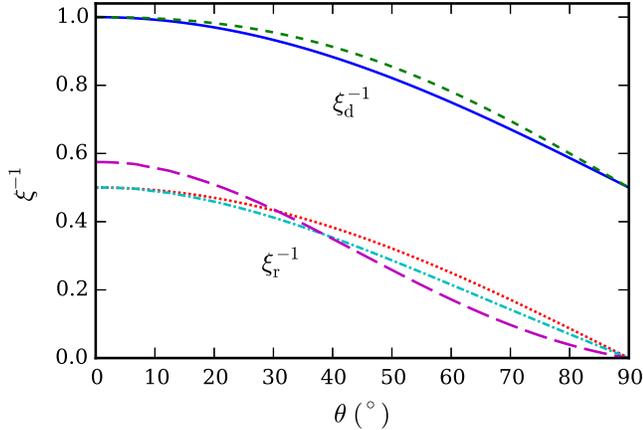}
\caption{Anisotropy factors, $\xi_\mathrm{d,r}$, as a function of the inclination angle, $\theta$, from numerical models of flat disks with different radiative distribution models (Section~\ref{sec:Radia}).
The star is Lambertian (solid, dotted, dash-dotted) or PES (short dashed, long dashed). The disk is Lambertian (dotted, dash-dotted) or follows an $H$-function (long dashed).
We compute the anisotropy factors with (dash-dotted, long dashed) or without (dotted) line-of-sight blocking of the disk by the star. 
\label{fig:compare}}
\end{figure}

We compare the anisotropy factors of models with a flat disk for different radiative distribution models (Section~\ref{sec:Radia}).
$\xi^{-1}_{\rm d}$ for a pure-electron-scattering neutron star is slightly larger than that of a Lambertian neutron star (Figure \ref{fig:compare}), because fewer photons reach the flat disk in the former than in the latter. 
In fact, $P=0.250$ in the case that the burster is Lambertian and $P=0.228$ in the case that the burster is pure-electron-scattering.

For the reflected burst emission, $\xi^{-1}_\mathrm{r}$ for a disk with an $H$-function displays the concentration of intensity towards smaller angles for this radiative distribution (Figure~\ref{fig:radia_transfer}). Furthermore, if the neutron star blocks part of the photons reflected off the disk, $\xi^{-1}_{\rm r}$ is a bit smaller when taking star-disk blocking into account, and the absolute difference is 0.036 at most (Figure \ref{fig:compare}).

\subsection{Gravitational Light Bending}
\label{sec:lb}

\begin{figure}
\centering
\includegraphics[width=1\linewidth]{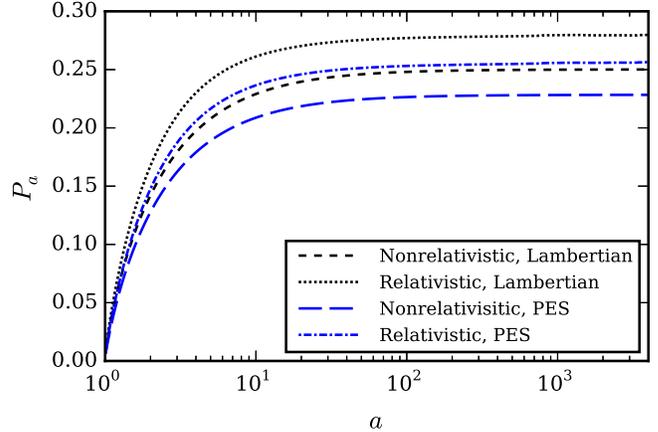}
\caption{Intrinsic reflection fraction $P_a$ considering the burst flux intercepted by the disk within radius $a=r/R_\star$ for a Lambertian and a PES neutron star, taking into account gravitational light bending. Most of the radiation falls on the inner disk ($a\lesssim10$).
  The difference in the total fraction $P$ caused by light bending is $11.9\%$ for the Lambertian case and $12.3\%$ for the PES case.
\label{fig:lb}}
\end{figure}

Due to its compactness, a neutron star's surface gravity is extremely strong, and light bending occurs in its vicinity. 
This increases the fraction of the burst flux that is intercepted by the disk.
We investigate the increase in $P$ by including light bending in the calculation for a flat disk.

We calculate a series of
photon paths starting at the stellar surface and initially making
an angle with the surface normal of $0\le\alpha\le\pi/2$. We employ
polar coordinates $u,\phi$, where $u$ is the inverse of the radial
distance from the star's center: $u\equiv r^{-1}$. Each
path is traced by using a fourth order Runge-Kutta scheme to solve the null-geodesic
in the Schwarzschild metric \citep[e.g.,][]{Hoyng2006book}: 
\begin{equation}
\frac{\mathrm{d}^{2}u}{\mathrm{d}\phi^{2}}+u=3mu^{2},\label{eq:geodesic}
\end{equation}
 with $m\equiv GM/c^{2}$ (half the Schwarschild radius), and
we use a neutron star mass of $M=1.4\,M_{\odot}$ and a radius of
$R_\star=10\,\mathrm{km}$. We take steps in $\phi$ of $10^{-3}$, which
gives a solution that is converged within $\sim10^{-13}$.

The photon paths are employed to trace where flux from the neutron star falls
on the disk (Figure \ref{fig:lb}). We find an increase of $\sim 12\%$ in $P$. 
This increase is small with respect to the large changes in reflection fraction
that we are interested in. Therefore, we will neglect relativistic effects in the
remainder of our study.

\subsection{Linearly Inclined and Concave Disks}

\begin{figure}
\centering
\includegraphics[width=1\linewidth]{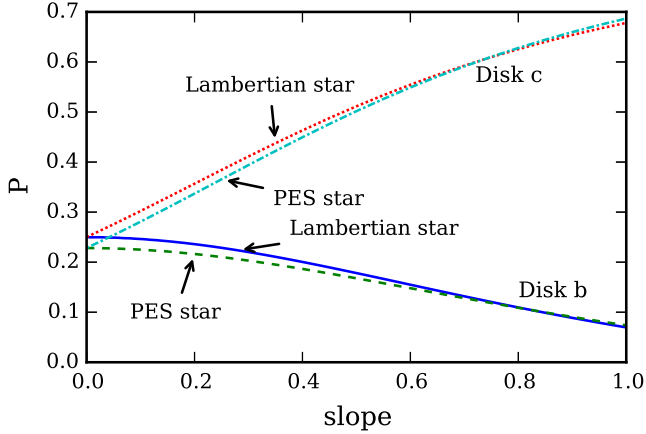}
\caption{Intrinsic reflection fraction, $P$, as a function of the linear slope of the radial height profile of the disk. Shown are disks with shapes b and c (Figure \ref{fig:disk}), both with a Lambertian and a pure-electron-scattering (PES) stellar atmosphere.
For the disk with shape b, part of the star is covered by the disk, so $P$ decreases as the slope of the disk height increases.
In contrast, the disk shape c, which has a triangular cross section, does not cover the star. For larger slopes, the disk subtends to larger solid angles, which increases $P$.
\label{fig:RF}}
\end{figure}

In this section, we explore the angular distribution of X-ray radiation for disks with other shapes than flat, and we consider linearly inclined and concave disks (Figure~\ref{fig:disk}).
A so-called $\alpha$-disk \citep{Shakura1973} with a typical viscosity parameter of
$\alpha=0.1$ \citep[e.g.,][]{King2007} is similar to our ``c'' shape with a slope of $0.01$.
We will, however, focus on much larger slopes, to investigate larger deviations
from the flat disk case.

We first study the intrinsic reflection fraction, which has a large influence on the anisotropy factors. Figure \ref{fig:RF} shows $P$ as a function of the slope of the height of the trapezoidal (shape b in Figure~\ref{fig:disk}) and triangular (shape c) disk. For shape b, $P$ descreases with slope, because a disk with a larger slope covers a larger part of the stellar surface. For disks with shape c, however, a larger slope means that more burst flux is intercepted by the disk surface: for a slope of $1.0$, the intrinsic reflection fraction has a large value of $P=0.65$. The difference in $P$ between models where the star emits as a Lambertian or a pure-electron-scattering atmosphere are small (Figure~\ref{fig:RF}), similar to what we saw for flat disks (Figure~\ref{fig:compare}).

\subsubsection{Flux Distribution Received by the Disk Surface}

\begin{figure}
\centering
\includegraphics[width=1\linewidth]{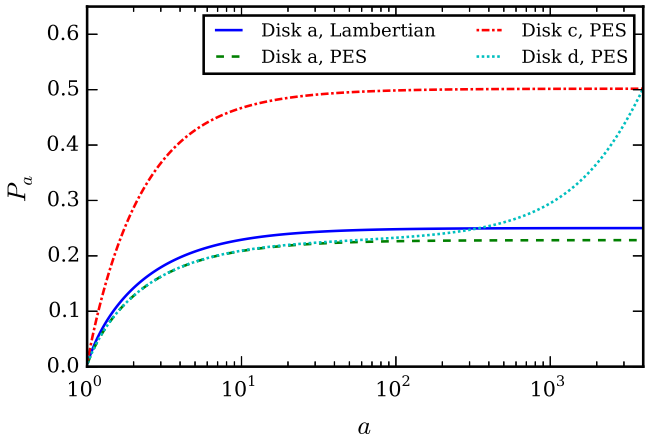}
\caption{Similar to Figure~\ref{fig:lb} for disk shapes a, c and d (Figure~\ref{fig:disk}).
  For all shapes we assume a PES stellar atmosphere, and for disk a we also include a model with a Lambertian star.
  Disk c, with a slope of $0.5$, subtends greater solid angles than a flat disk as seen from the stellar center, thereby accumulating more flux on its surface. Disk d curves up, collecting a substantial amount of flux at larger radii.
\label{fig:accu_disk2}}
\end{figure}

For flat disks, we found that most of the burst flux received by the disk falls on the inner disk (Table~\ref{tab:1}). 
Assuming a PES star, 90\% of the radiation falls within $8.6R_\star$ (Figure \ref{fig:accu_disk2}). Also for linearly inclined disks (Figure~\ref{fig:disk}c) most of the flux falls in a similarly small region (Figure \ref{fig:accu_disk2}). 
However, for a concave disk of which the outer part goes up sharply (Figure~\ref{fig:disk}d), half of the flux falls on the region outside of $100R_\star$. 

At the outer radius of a flat disk around a Lambertian star, we find a total intrinsic reflection
fraction of $P=0.25$, whereas for disk d it is $P=0.5$, which means half of the photons emitted by the neutron star irradiate the disk.

\subsubsection{Anisotropy Factors}

\begin{figure}
\centering
\includegraphics[width=1\linewidth]{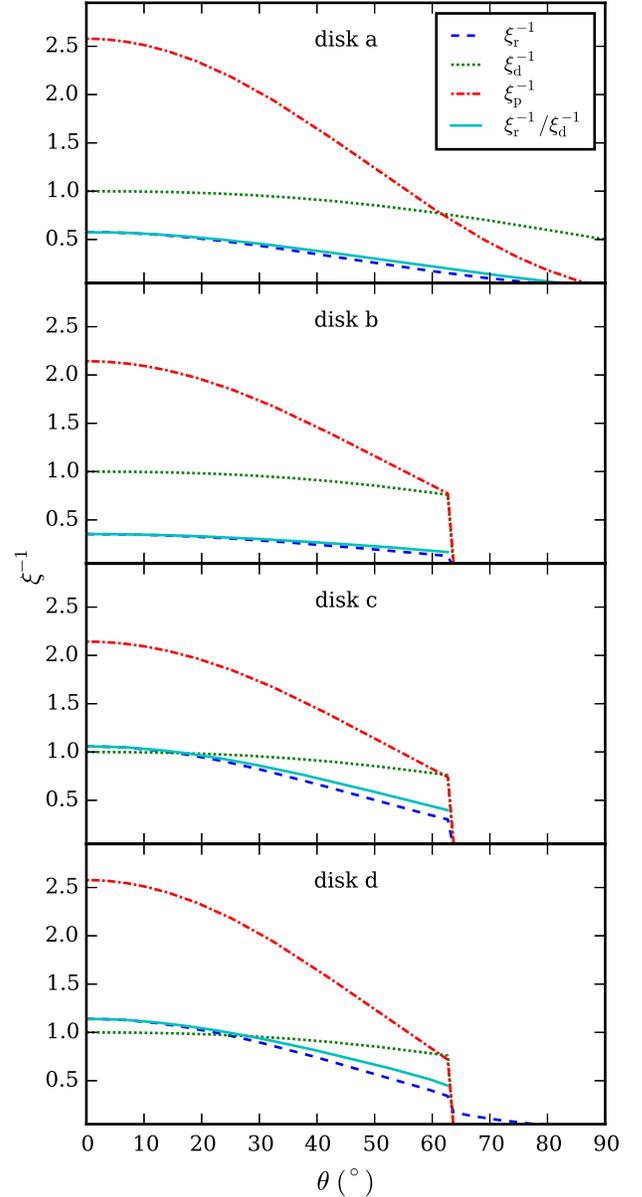}
  \caption{\label{fig:xi_all}
  Anisotropy factors, $\xi_{\mathrm{d,r,p}}^{-1}$,
    and reflection fraction, $\xi^{-1}_\mathrm{r}/\xi^{-1}_\mathrm{d}$, as observed under inclination angle $\theta$ for different disk
    shapes (see Figure 1). Disk a is flat, disks b and c have a linearly
    increasing height with slope $0.5$, and disk d has a quadratically
    increasing height that is normalized such that at the outer edge it
    is the same as for disks b and c.
    }
\end{figure}

We calculate the anisotropy factors and observed reflection fractions for all disk shapes in Figure~\ref{fig:disk} (Figure~\ref{fig:xi_all}), where we use pure electron scattering for the neutron star and an $H$-function for the disk (Section \ref{sec:Radia}). The anisotropy factors of the direct burst flux, $\xi^{-1}_\mathrm{d}$, is the same for all disk shapes, as long as the view of the star is not blocked by the disk. For the chosen slopes of disks b, c, and d, the center of the star and the outer edge of the disk line up with the observer's line of sight for an inclination of $\theta\simeq63.4^\circ$. Because the outer radius of the disk is much larger than the radius of the star, a small increase in $\theta$ of only $\sim 0.01^\circ$ completely hides the star from view, and reduces $\xi^{-1}_\mathrm{d}$ to $0$.

There is some variation in $\xi^{-1}_\mathrm{r}$ and $\xi^{-1}_\mathrm{p}$ as a function of disk shape. Most notably, $\xi^{-1}_\mathrm{r}$ is about twice as large for disks c and d as it is for a flat disk. When the direct view of the star is blocked by the disk, disk d's upturned outer edge still allows some reflected burst flux to reach the observer ($\theta\gtrsim 63.4^\circ$ in Figure~\ref{fig:xi_all}d).

The observed reflection fraction, $\xi^{-1}_\mathrm{r}/\xi^{-1}_\mathrm{d}$, follows similar trends as $\xi^{-1}_\mathrm{r}$. For small inclination angles it even exceeds unity for shape c and d: most of the observed burst radiation is reflected off the disk. We saw that for disk c with the same slope, the intrinsic reflection fraction is $P=0.5$ (Figure~\ref{fig:RF}). Therefore, it is due to the anisotropies of the system that the observed reflection fraction exceeds unity.

For the disk shapes considered so far, the observed reflection fraction peaks at $\theta=0^\circ$. We can easily understand this for our analytic model, as $\xi^{-1}_{\rm r}$ drops faster with $\theta$ than $\xi^{-1}_{\rm d}$, because $\xi^{-1}_{\rm r}$ falls off as $\cos\theta$ whereas $\xi^{-1}_{\rm d}$ decreases like $1+\cos\theta$. The maximum is only a little bit larger than $1$. We investigate the maximum observed reflection fraction $\xi^{-1}_{\rm r}/\xi^{-1}_{\rm d}$ for any disk shape. 

For all shapes, the entire neutron star is visible at $\theta=0^\circ$: $\xi^{-1}_{\rm d}=1$. This reduces the problem to finding the maximum $\xi^{-1}_{\rm r}$.
The analytic model shows that $\xi^{-1}_{\rm r}$ consists of two factors: a factor from the radiation distribution model and $P$ (Equation~\ref{eqn:r}). The former is maximal for a flat disk, where every part of the disk surface is oriented the same. For a flat $H$-function-like disk this factor is $\xi^{-1}_{\rm r}/P=2.52$ (Section~\ref{sec:radiation_flat}, Figure~\ref{fig:xi_all}a). For other disk shapes, therefore, $\xi^{-1}_{\rm r} \leq 2.52 P$.

The physical limit for $P$ is $1$, which means all of the burst emission is intercepted by the disk. Although not $100\%$, we found that a disk with shape c can intercept the majority of the burst flux, depending on the slope (Figure~\ref{fig:RF}). For shape b, however, $P$ decreases for increased slopes, because a larger part of the star is blocked by the inner disk. Therefore, we conclude that for any concave disk shape, $\xi^{-1}_{\rm r} / \xi^{-1}_{\rm d}$ could never exceed $2.52$. An exception is the special case of partial blocking, which we discuss next.



\subsubsection{Partial Blocking by a Steep Inner Disk}
\label{sec:partial_blocking}

\begin{figure}
\centering
\includegraphics[width=1\linewidth]{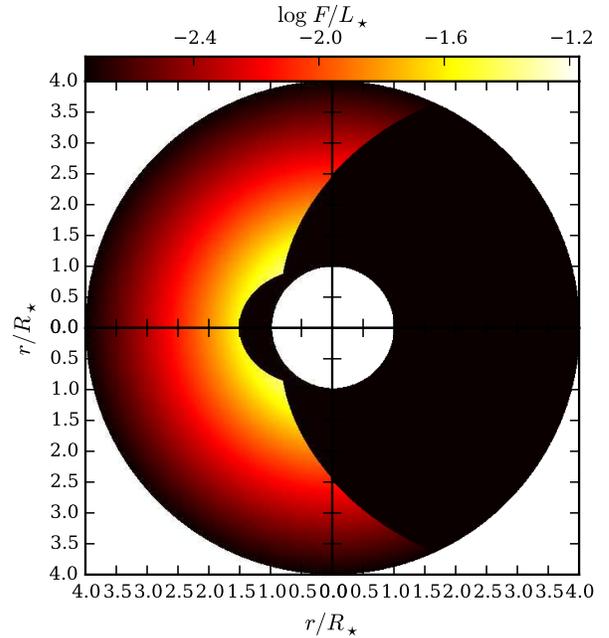}
\caption{Distribution of burst flux received by the disk. The disk has shape d (see Figure \ref{fig:disk}),
  a radial height profile of $h/R_\star=0.57735(r/R_\star-1)^{1.5}$,
  an outer radius of $4R_\star$, and is $H$-function-like; the star radiates as a pure-electron-scattering surface.
  Black areas are blocked from view by the disk and the star for an inclination angle of $58^\circ$. The ``shadow''
  of the disk partially falls across the neutron star: the observer sees a large reflection fraction.
\label{fig:pattern}}
\end{figure}

Consider a disk where the inner part steeply increases in height and the outer part is relative flat. Burst photons only hit the inner disk, which shields the outer part of the disk. For our purposes, such a geometry behaves identically to a system with a small steep disk. We employ disks with an outer radius of $4R_\star$. Figure~\ref{fig:pattern} illustrates where the burst flux falls on a disk with shape d (Figure \ref{fig:disk}), and which parts of the disk are blocked from the observer's view. For particular inclination angles, a substantial part of the star is blocked as well, which produces a large observed reflection fraction.

\begin{figure}
\centering
\includegraphics[width=1\linewidth]{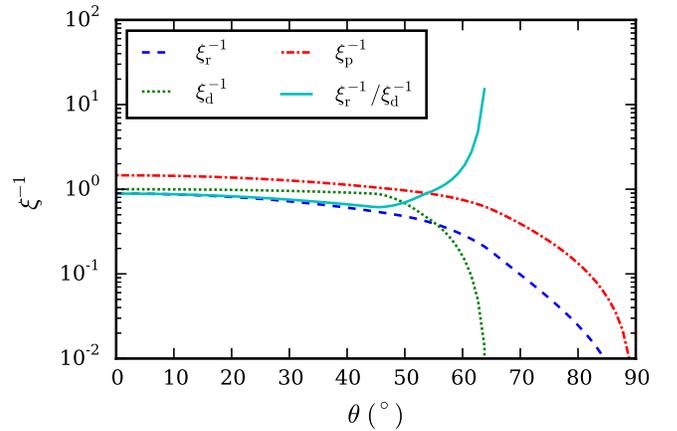}
\caption{Anisotropy factors, $\xi^{-1}_\mathrm{d,r,p}$, of a shape c disk with an outer radius of $4R_\star$ and a slope of $1$. This disk has the same outer radius and height as shown in Fig. \ref{fig:pattern}, which has shape d. The neutron star is pure-electron-scattering and the disk is $H$-function-like. The steeply rising inner disk leads to a large observed reflection fraction, $\xi^{-1}_\mathrm{r}/\xi^{-1}_\mathrm{d}$.
\label{fig:ani_R4_1}}
\end{figure}
Similarly, large reflection fractions can be obtained for shape c. There is a substantial range of inclination angles for which the star is partially blocked by a small linearly inclined disk (Figure \ref{fig:ani_R4_1}). At around $\theta = 60^\circ$, the direct burst flux reduces quickly as the star is being blocked by the disk, whereas the reflected burst flux changes more slowly, resulting in observed reflection fractions as large as $\xi^{-1}_\mathrm{r}/\xi^{-1}_\mathrm{d}\sim 10$.

\section{Discussion}

Anisotropy factors were calculated for different accretion disk shapes.
We discuss how our simulations compare to previous models of flat
disks, and we investigate the impact on observable properties, including
the reflection fraction and the $\alpha$-parameter. Furthermore,
we consider issues where anisotropies may play an important role,
such as mass-radius measurements.

\subsection{Comparison to Previous Models of Flat Disks}

\begin{figure}
\centering
  \includegraphics[width=1\linewidth]{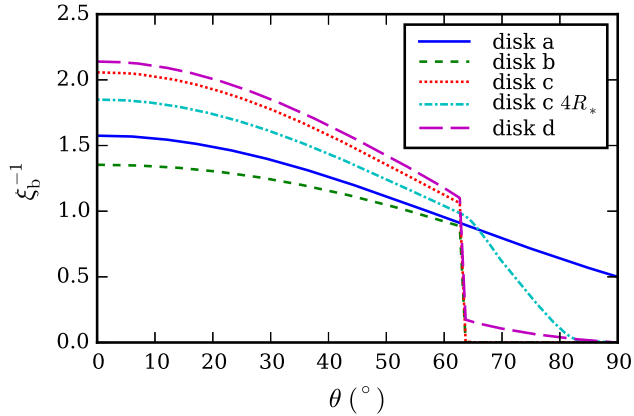}
  \caption{\label{fig:xi_b}
  Total burst anisotropy factor, $\xi_{\mathrm{b}}^{-1}$, as a
    function of the inclination angle, $\theta$, for the different disk shapes
    presented in Figures~\ref{fig:xi_all} and \ref{fig:ani_R4_1}.}
\end{figure}
For a flat disk and Lambertian surfaces our numerical model closely
reproduces the analytic result of \citet{fujimoto88apj}. We also
consider the refinements to the model introduced by \citet{Lapidus1985}:
more realistic radiation distribution models by using a pure-electron-scattering
atmosphere for the neutron star and an $H$-function for emission from
the disk, as well as blocking of part of the line of sight by a ``shadow''
of the star and disk. We compare the results of our model ``a'' from
Figure~\ref{fig:xi_b} to their model 3 \citep[Figure 2 in][]{Lapidus1985}.
Both models find the fraction of the burst luminosity that is intercepted
by the disk to be $P=0.23$, and at $\theta=90^{\circ}$ both find
$\xi_{\mathrm{b}}^{-1}=0.5$ (Figure~\ref{fig:xi_b}; $\xi_{\mathrm{b}}^{-1}$
is called $F_{\mathrm{syst}}$ in \citealt{Lapidus1985}). At lower
inclinations, however, small differences are present. Our $\xi_{\mathrm{b}}^{-1}$
is up to $6\%$ smaller than the Lambertian case for $\theta>38^{\circ}$,
and larger at lower inclination, with $\xi_{\mathrm{b}}^{-1}=1.58$
for $\theta=0$. This behavior is driven by the shadow blocking part
of the burst flux for large $\theta$, whereas electron scattering
and the $H$-function distribute the flux more to small $\theta$ compared
to Lambertian surfaces (Figure~\ref{fig:radia_transfer}). In constrast, \citet{Lapidus1985}
find $\xi_{\mathrm{b}}^{-1}$ to be slightly larger than the Lambertian
case for $\theta>68^{\circ}$, and smaller at lower $\theta$, with
$\xi_{\mathrm{b}}^{-1}=1.39$ for $\theta=0$. \citet{Lapidus1985}
do not provide sufficient detail for us to explain the discrepancy,
but it is likely related to differences in the implementations of
blocking or the $H$-function. \citet{fujimoto88apj} noted that the
results of \citet{Lapidus1985} lead to inconsistencies in mass-radius
measurements, whereas our results are consistent (Section~\ref{sub:The--parameter}).
The differences between the models are, however, smaller than the
large changes in anisotropy that we are searching for. 

Similarly, we ignore the effect of light bending, which increases
the luminosity intercepted by the disk by only $\sim12\%$ (Section~\ref{sec:lb};
see also \citealt{Lapidus1985}).

\subsection{Intrinsic Reflection Fraction for Concave Disks}

We consider concave disks, where the height of the disk increases
monotonically with radius. Standard $\alpha$-disk theory \citep{Shakura1973}
predicts that for a typical viscosity parameter of $\alpha=0.1$,
the slope of the height as a function of radius, $r/R_\star$, is small:
$\sim10^{-2}$. We consider much larger slopes ($0.5$ for disk b
and c in Figure~\ref{fig:xi_all}) in order to investigate large
deviations from a flat geometry. Such large deviations could potentially
be induced by the irradiation of the disk by the X-ray burst \citep{Ballantyne2005}. 

With the exception of shape b, where part of the stellar surface is
covered by the disk, concave disks have substantially larger instrinsic
reflection fractions. Whereas flat disks capture $P=0.23$ of the
star's luminosity, shape c intercepts as much as $P=0.68$ (slope
of $1.0$, see Figure~\ref{fig:RF}). Shape d has $P$ values as large as shape
c, but more of the star's flux is intercepted at larger radii, due
to the steepening of the height profile further away from the star
(Figure~\ref{fig:disk}).

\subsubsection{Reflection Fractions for AGN}

It is interesting to compare our results to those obtained for X-ray
reflection from accreting black holes, including AGN \citep[e.g.,][]{Fabian2010review}.
In this case the disk is illuminated by a corona above the accretion
disk. This can lead to reflection fractions as high as $P\simeq0.84$,
as light emitted downwards is reflected upwards \citep{Blackman1999}.
Furthermore, strong light bending near the inner radius of the disk
causes a large fraction of the coronal emission to be intercepted
and reflected by the disk: intrinsic reflection fractions as large
as $P\simeq0.97$ are predicted, depending on the height of the corona
above the disk \citep{Dauser2014}. For accretion onto neutron stars,
the gravitational potential at the inner disk is less strong, increasing
$P$ by only $\sim12\%$ \citep[see also][]{Lapidus1985}, and downward
emission from the bottom of the star cannot be reflected upwards. 

\citet{Fabian2002} suggest a disk geometry where the X-ray source
is embedded in a ring structure. For all but very small inclination
angles, the source is hidden from view, but reflection can still be
observed. This is analogous to our situation of a neutron star embedded
in a disk with a fast rising height (Figure~\ref{fig:ani_R4_1}).

Concave disks may change the shape of spectral features. For example,
the prominent Fe~K$\alpha$ line around $6.4\,\mathrm{keV}$ undergoes
substantial Doppler broadening at the inner disk, whereas it remains
narrow when originating from the outer disk. For a concave disk the
contribution to the line from the outer disk is larger, resulting
in a relatively small broad component \citep{Hartnoll2000}. We have
not considered the changes of reflection features in X-ray burst spectra.
The present data with reflection of two superbursts prefer a broad
line, but are of insufficient quality to separate broad and narrow
components.

\subsection{Observable Quantities}

Because of substantial uncertainties in the distance to burst sources,
the accretion composition, and the inclination angle, it is challenging
to derive anisotropy factors from observations \citep[e.g.,][]{Heger2007}.
Taking ratios of flux components has an advantage, because it removes
the distance dependence. \citet{fujimoto88apj} and \citet{Lapidus1985}
compared their results with observed values of the $\alpha$-parameter.
Because we consider the direct and reflected burst flux separately,
we additionally discuss the observed reflection fraction.

\subsubsection{Observed Reflection Fraction}

Depending on the inclination angle, the observed reflection fraction
can be substantially larger than $P$. For shapes c and d, it can
even be slightly larger than unity: the majority of the detected burst
flux is reflected off the disk (Figure~\ref{fig:xi_all}). Even larger
reflection fractions of up to $\sim10$ are only expected in case
the disk height increases substantially at relatively small radii,
such that part of the star is blocked from the line of sight (Figure~\ref{fig:ani_R4_1}). 

X-ray reflection has only been detected during two events, both observed
with the proportional counter array on the \emph{Rossi X-ray Timing
Explorer}: the 1999 superburst from 4U~1820--30 \citep{Ballantyne2004}
and the 2001 superburst from 4U~1636--536 \citep{Keek2014sb2}. Around
the time that the flux peaked, reflection fractions of $\sim0.2$
and $0.7$ are observed, respectively. Whereas the former value can
easily be accomodated with a flat disk, the latter value is slightly
larger than the maximum value predicted for a flat geometry. In the
tail of both superbursts, the reflection fraction is observed to increase
substantially to $\sim3$ and $6$, respectively. Values this high
can only be produced by concave disks with partial blocking of the
star from the line of sight. This evolution of the reflection fraction
suggests that the geometry of the disk changed from flat to concave
under influence of intense irradiation by the superbursts. X-ray heating
may cause the inner disk to puff up \citep{Ballantyne2005}, which
would be consistent with a geometry that can hide part of the star
from view. The burst flux and the reflection signal are, however,
weaker in the tail. \citet{Keek2015sb3} showed that for 4U~1636--536
an alternative interpretation of the spectra exists, where the reflection
fraction is unchanged from the value at the peak. That interpretation,
however, has issues as well. New observations with future instrumentation
such as \emph{NICER} are required to measure the evolution of the
reflection fraction during bursts with greater confidence.

\subsubsection{The $\alpha$-Parameter}
\label{sub:The--parameter}

\begin{figure}
\centering
\includegraphics[width=1\linewidth]{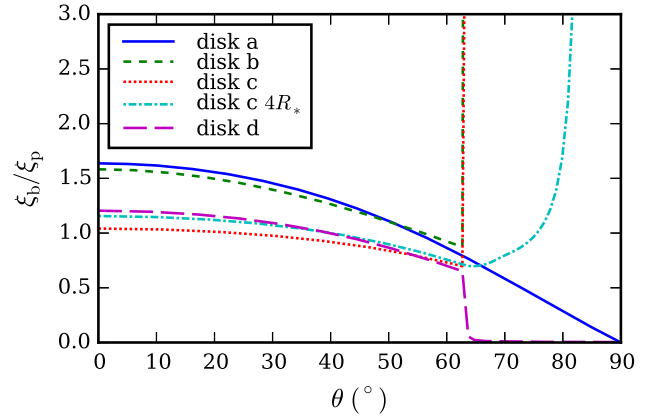}
\caption{\label{fig:alpha}$\xi_{\mathrm{b}}/\xi_{\mathrm{p}}$ for different
disk shapes, as presented in Figures~\ref{fig:xi_all},~\ref{fig:xi_b}.
The observed $\alpha$ parameter differs from the intrinsic value
by this factor.}
\end{figure}
The $\alpha$-parameter is the ratio of the persistent fluence between
two subsequent bursts to the fluence in one burst. It is generally
considered to be a measure of how much of the accreted hydrogen and
helium burns in the burst vs. stably in between bursts. The material
that is burned in a stable manner has only a small contribution to
the persistent fluence, as the gravitational potential energy that
is liberated by the accretion process is much larger than the energy
gained from nuclear burning. Its effect is, however, noticable in
the burst fluence, since only part of the accreted fuel is burned
in the burst. Anisotropic emission, therefore, changes the observed
$\alpha$-parameter from the intrinsic value by a factor $\xi_{\mathrm{b}}/\xi_{\mathrm{p}}$
\citep{fujimoto88apj}. To compare $\alpha$ between sources with
different inclination angles or with theoretical predictions, the
observed $\alpha$ values must be divided by this factor.

For flat Lambertian disks, the observed $\alpha$ is larger than the
intrinsic value by up to $33\%$ for small inclination angles ($\theta<60^{\circ}$),
and smaller for larger angles, as $\xi_{\mathrm{p}}^{-1}$ becomes
vanishingly small for $\theta$ close to $90^{\circ}$ \citep{fujimoto88apj}.
Our numerical model of a flat disk with blocking and improved radiation
models exhibits stronger changes in $\alpha$: the observed value
is larger by up to $64\%$ (Figure~\ref{fig:alpha}). The 
disk with shape b has similar deviations. For disks with shapes
c and d, however, the observed $\alpha$ is closer to the intrinsic
value: for our choice of disk shapes, at $\theta=0^{\circ}$ the observed
value is larger by at most $20\%$. At large inclination angles the
star and inner disk are blocked from the line-of-sight. This produces
either very small or very large values of $\alpha$ in the simulations,
neither of which are observable in practice when both burst and persistent
flux are almost fully blocked. The only exception is the case where
the height of the inner disk increases fast (Figure~\ref{fig:ani_R4_1}), such that
in a relatively large range of $\theta$ the view of the star is partially
blocked. This situation increases the observed $\alpha$ parameter
by up to a factor $\sim6$ (disk c with $4\,R_\star$ in Figure~\ref{fig:alpha}).

\citet{Heger2007} estimated the ratio of the persistent and burst
anisotropy factors for GS~1826--24 by comparing numerical models
to burst observations, and find $\xi_{\mathrm{b}}/\xi_{\mathrm{p}}=0.65$.
This value can be reproduced with any of the disk shapes that we consider.
For disk shape a, this value is reached at $\theta=68^{\circ}$, and
for our disk shapes c and d, it is just reached at approximately $\theta=63^{\circ}$
where blocking of the star by the disk starts.

Considering a large
sample of bursts from many sources, \citet{Galloway2008catalog} found
only long bursts when $\alpha<60$, whereas short bursts occur with
larger values of $\alpha$. For both groups of bursts, there is considerable
spread in $\alpha$. Furthermore, \citet{Zand2003} compiled a list
of $\alpha$ values from various studies, and found that $\alpha$
is substantially higher for superbursting sources than for bursting
sources that lack superbursts, which could be explained by the substantial
stable burning that may be required to produce the carbon fuel for
superbursts. Part of the spread in $\alpha$ may, therefore, be related
to different accretion compositions or burning behavior, whereas another
part will be due to differences in the anisotropy factors of the sources
in the sample. 

The $\alpha$ parameter has been employed to measure the ratio of
the neutron star radius and mass \citep{fujimoto88apj}. The main
source of uncertainty is $\xi_{\mathrm{b}}/\xi_{\mathrm{p}}$, which
depends on the poorly constrained inclination angle. For two sources,
the mass-radius constraints were calculated using $\xi_{\mathrm{b}}/\xi_{\mathrm{p}}$
from \citet{fujimoto88apj} and \citet{Lapidus1985}. The results
are roughly consistent, except with \citet{Lapidus1985} for the high
inclination source EXO~0748-676, which yielded an exceptionally large
radius of $\gtrsim20\,\mathrm{km}$. Using our values of $\xi_{\mathrm{b}}/\xi_{\mathrm{p}}$
for flat disk a (Figure~\ref{fig:alpha}) and all other parameters
from \citet{fujimoto88apj}, we find neutron star parameters similar
to \citet{fujimoto88apj}, which are consistent between large and
small inclination angle ($R\simeq(5.5-11)(M/1.4\,M_{\odot})\,\mathrm{km}$).

\subsection{Applications of Anisotropies and Reflection}

With the present observations, it is challenging to constrain the
anisotropy factors. Nevertheless, anisotropy and reflection have important
consequences for the interpretation of burst observations. We discuss
several topics where they may play a role.

\subsubsection{Distance and Mass-Radius Measurements}

The peak flux of bursts with photospheric radius expansion (PRE) is employed
as a measure of the Eddington limit, which is used for distance determination
\citep[e.g.,][]{Kuulkers2003} and to constrain the neutron star compactness
\citep[e.g.,][]{Ozel2006,Galloway2008pre}. These measurements are
all biased by $\xi_{\mathrm{b}}^{-1}$, which is generally not taken
into account. For example, different values of the Eddington limit
have been observed for 4U~1636--536 \citep{Galloway2006}. It has
been suggested to be due to different photospheric composition, but
changing anisotropy factors may produce a similar effect if the disk
geometry evolves with the persistent flux. Furthermore, the observed
spread in the Eddington luminosity between different sources \citep{Kuulkers2003}
will in part be due to different anisotropies.

A different method to constrain the neutron star compactness uses
detailed models of neutron star atmosphere spectra, parametrizing
deviations from a blackbody as a color-correction factor \citep{Suleimanov2010}.
This method was found to work well when the persistent flux is low
and in the hard state, but not at high flux in the soft state \citep{Kajava2014}.
In the soft state the observed burst spectra generally do not follow
the predicted evolution of the color correction, and the behavior
of the normalization of the burst spectra is different from the hard
state. \citet{Kajava2014} speculate that the accretion geometry is
different in the two states. In the soft state, a spreading layer
between the disk and the neutron star surface could reprocess a substantial
part of the burst flux, modifying its spectrum. Alternatively, the
different accretion geometry may produce an increased reflection fraction,
such that reflection accounts for a larger part of the burst spectrum.
As the reflection spectrum is reprocessed by the disk \citep{Ballantyne2004models},
this may explain why it does not conform to the spectral evolution
predicted for neutron star atmospheres.

\subsubsection{Variability in Persistent and Burst Flux}

If the accretion geometry evolves as a function of the persistent
flux \citep[e.g.,][]{Done2005review}, anisotropies may be different
for bursts in different persistent states for a single source. Moreover,
recently indications have been found that anisotropies may even evolve
during a single X-ray burst.

\citet{Worpel2013} found that the observed persistent flux briefly
increases during bursts, which they measure as a multiplicative factor,
$f_{\mathrm{a}}$ \citep[see also][]{Zand2013,Worpel2015}. The anisotropy
parameter $\xi_{\mathrm{p}}^{-1}$ has a similar effect on the observed
flux as $f_{\mathrm{a}}$. Its variation for different shapes is,
however, smaller than the observed $f_{\mathrm{a}}$ factors (Figure~\ref{fig:xi_all}).
Therefore, it seems unlikely that $f_{\mathrm{a}}$ is caused by an
evolving $\xi_{\mathrm{p}}^{-1}$. This is confirmed by the 2001 superburst
from 4U~1636--536, where the observed increase in persistent emission
was not directly correlated with the evolution of the reflection fraction
\citep{Keek2014sb1,Keek2014sb2}. Poynting-Robertson drag has been
suggested to cause a temporary increase of the persistent flux \citep{Worpel2013,Ballantyne2005}.
This would evacuate the inner disk, and potentially change the anisotropy
factors during a burst. 

In a small number of bursts with exceptionally large radius expansion
(``superexpansion''), strong variability is observed in the tail
of the light curve \citep{Zand2011,Degenaar2013}. The flux both exceeds
and dips below the burst's cooling trend, which has been explained
as alternate blocking by and reflection off surrounding material.
This could be an example of anisotropies from an accretion environment
that is not symmetric with respect to the rotation axis of the disk.

\section{Conclusions \& Outlook}

We have used numerical models to calculate the anisotropy of X-ray
emission from LMXBs with neutron stars that produce Type I X-ray bursts.
We separately present the anisotropy factors of the directly observed
burst flux and the flux reflected off the accretion disk, as well
as of the persistent flux. Our models account for different disk geometries,
including concave disks. The latter produce observed reflection fractions
much larger than unity, when part of the star is blocked from view
by the disk. Such large reflection fractions have recently been inferred
for the tail of two superbursts. A strongly concave disk, however,
could completely block our view of bursts from high-inclination sources,
but this is not the case for EXO~0748-676 \citep{0748:parmar86apj}.
The situation is likely complex, where the shape of the accretion
disk may change between spectral states of the source, and even evolve
during an X-ray burst. This complicates making quantitative predictions
of the anisotropy factors for specific LMXBs or bursts. We discussed
the qualitative effects of the anisotropies on observed quantities,
including the reflection fraction and the $\alpha$-parameter.

Our models can be improved with a physics-based underpinning of the
assumed accretion geometries, including a spreading layer and a gap
between the neutron star and the inner disk. As multiple processes
may be of importance during X-ray bursts \citep{Ballantyne2005},
detailed numerical models are required to accurately capture the behavior
of accretion disks that are irradiated by bursts.

Important observational constraints on the anisotropy factors will
be provided by new X-ray instrumentation, as \emph{NICER} (scheduled
for launch in 2016) and \emph{Athena} (planned for launch in the late
2020s) will be able to measure reflection features during bright X-ray
bursts (Wolf et al. 2016 in prep.). In the mean time, better measurements
of the $\alpha$-parameter can be obtained by employing the Multi-Instrument
Burst Archive \citep[MINBAR;][]{Keek2010}. Furthermore, the anisotropy
parameters depend strongly on the inclination angle, which is currently
poorly constrained. Simultaneous X-ray and optical observations may
improve this situation \citep[e.g.,][]{MunozDariaz2008}.

\acknowledgments{The authors thank D.~Ballantyne and T.~Strohmayer for helpful
discussions, and acknowledge the Center for Relativistic Astrophysics at Georgia
Institute of Technology, where this study was initiated. CCH is supported by the Undergraduate Education Office of Jilin University in Changchun, China, which also supports this publication. LK is supported by NASA 
under award number NNG06EO90A. LK thanks the International Space Science Institute
in Bern, Switzerland for hosting an International Team on X-ray bursts.
}

\bibliographystyle{apj}
\bibliography{references}

\appendix

\section{Reflection Fraction of a Disk with Finite Outer Radius} 
\label{app:a}
In Section \ref{sec:Ana}, we have computed analytically the intrinsic reflection fraction, $ P $, of a disk with infinite outer radius. Here we compute $ P_a $ of a disk with an outer radius of $ r=a R_\star $, where $ R_\star $ is the stellar radius.

We use the same coordinate systems and parameter definitions as in Section \ref{sec:Ana} (Figure \ref{fig:coordinate}).
$ \delta $ as a function of $ \varphi $ is given by
\begin{equation} \label{tandelta}
\tan \delta 
= \frac{\sin \varphi}{\cos \theta_0} \left ( \sin \theta_0 - \frac{1}{a \sin \beta} \right ) ,
\end{equation}
where $ \beta $ is the azimuthal angle in the $ o_0x_0y_0z_0 $ coordinate system of the point on the disk edge that $ oE $ points to (Figure \ref{fig:coordinate}). Since $ \delta > 0 $, Equation (\ref{tandelta}) implies
\begin{equation} \label{sintheta}
\sin \theta_0 \geq \frac{1}{a \sin \beta},
\end{equation}
where
\begin{equation} \label{sinalpha}
\sin \beta = \frac{\tan \varphi}{\sqrt{\tan^{2} \varphi+\cos^{2} \theta_0}}.
\end{equation}
Substituting Equation (\ref{sinalpha}) into Equation (\ref{tandelta}), we find
\begin{equation} \label{Eqn:tandelta}
\tan \delta = \frac{\sin \varphi}{\cos \theta_0} \left ( \sin \theta_0 - \frac{\sqrt{\tan^{2} \varphi+\cos^{2} \theta_0}}{a \sin \varphi} \right ) \label{eqa:8}.
\end{equation}
Combining Equation (\ref{sintheta}) and Equation (\ref{sinalpha}), we obtain the range of $ \tan\varphi $
\begin{equation}\label{sinphi}
\tan^{2} \varphi \geq \frac{\cos^{2} \theta_0}{\left (a \sin \theta_0 \right )^{2}-1}.
\end{equation}
Similarly, we can establish from Equation (\ref{sintheta}) that
\begin{equation}\label{sint}
\sin \theta_0 \geq \frac{1}{a}.
\end{equation}
For one element on the stellar surface with polar angle $\theta_0$, the reflection fraction is given by
\begin{eqnarray}
P_a \left ( \theta_0 \right ) &=& \frac{1}{\pi} ~ \int_{\arctan \left ( \frac{\cos^{2} \theta_0}{\left (a \sin \theta_0 \right )^{2}-1} \right )}^{\pi - \arctan \left ( \frac{\cos^{2} \theta_0}{\left (a \sin \theta_0 \right )^{2}-1} \right )} \mathrm{d} \varphi \int_{\frac{\pi}{2}-\delta}^{\frac{\pi}{2}} \mathrm{d} \alpha \cos \alpha \sin \alpha \nonumber \\
&=& \frac{1}{2\pi} ~ \int_{\arctan \left ( \frac{\cos^{2} \theta_0}{\left (a \sin \theta_0 \right )^{2}-1} \right )}^{\pi - \arctan \left ( \frac{\cos^{2} \theta_0}{\left (a \sin \theta_0 \right )^{2}-1} \right )} \mathrm{d} \varphi \sin^{2} \delta.
\end{eqnarray}
So the total intrinsic reflection fraction is
\begin{eqnarray}
P_a &=& \frac{1}{2\pi} \int_{\arcsin\frac{1}{a}}^{\frac{\pi}{2}} \mathrm{d}\theta_0 \int_{0}^{2\pi} \mathrm{d}\phi_0 \sin \theta_0 P_a \left ( \theta_0 \right ) \nonumber \\
&=& \frac{1}{2\pi} ~ \int_{\arcsin\frac{1}{a}}^{\frac{\pi}{2}} \mathrm{d}\theta_0 ~ \int_{\arctan \left ( \frac{\cos^{2} \theta_0}{\left (a \sin \theta_0 \right )^{2}-1} \right )}^{\pi - \arctan \left ( \frac{\cos^{2} \theta_0}{\left (a \sin \theta_0 \right )^{2}-1} \right )} \mathrm{d} \varphi \frac{\tan^{2}\delta}{\tan^{2}\delta + 1} \sin \theta_0.  \label{RF}
\end{eqnarray}
We compute $ P_a $ for several values of $ a $ (Table \ref{tab:1}).

\begin{table}
\centering
\caption{Fraction of burst photons that irradiate a flat disk, $ P_a $, for several values of outer disk radius $a$. \label{tab:1}}
\renewcommand{\thefootnote}{$\star$} 
\begin{tabular}{cccccc}
	\hline $ a^\dagger $ & 4 & 40 & 400 & 4000 & $\infty$ \\
	$P_a$ & 0.1978386 & 0.2446961 & 0.2494695 & 0.2499469 & 0.25 \\ 
	\hline
\end{tabular}

$ ^\dagger $ The outer radius of the disk is $ r=a R_\star $
\end{table}

\end{document}